\documentclass[prl,twocolumn,superscriptaddress,floatfix]{revtex4}
\usepackage{amsfonts}
\usepackage{amsmath}
\usepackage{amssymb}
\usepackage{graphicx}

\newcommand{\BE}{\begin{equation}}
\newcommand{\BEA}{\begin{eqnarray}}
\newcommand{\EE}{\end{equation}}
\newcommand{\EEA}{\end{eqnarray}}

\newcommand\Tr{\mathrm{Tr}\,}
\newcommand\R\rangle
\renewcommand\L\langle
\renewcommand\r\right
\renewcommand\l\left

\newcommand\e{\mathrm{e}}
\newcommand\ev{\mathbf{e}}
\newcommand\rv{\mathbf{r}}

\newcommand\Kv{\mathbf{K}}
\newcommand\kv{\mathbf{k}}
\newcommand\Qv{\mathbf{Q}}

\newcommand\g{\gamma}
\newcommand\s{\sigma}
\newcommand\up{\uparrow}
\newcommand\dn{\downarrow}

\begin{document}
\title{Correlation-induced triplet superconductivity on the graphene lattice}
\author{Peyman Sahebsara}
\affiliation{D\'{e}partement de physique and Regroupement qu\'{e}b\'{e}cois sur les mat\'{e}riaux de pointe, Universit\'{e} de Sherbrooke, Sherbrooke, Qu\'{e}bec, Canada, J1K 2R1}
\affiliation{Department of Physics, Isfahan University of Technology, Isfahan 84156-83111, Iran}
\author{David S\'{e}n\'{e}chal}
\affiliation{D\'{e}partement de physique and Regroupement qu\'{e}b\'{e}cois sur les mat\'{e}riaux de pointe, Universit\'{e} de Sherbrooke, Sherbrooke, Qu\'{e}bec, Canada, J1K 2R1}
\date{\today}

\begin{abstract}
We investigate the possibility of superconductivity on the graphene lattice within the repulsive Hubbard model using the variational cluster approximation (VCA). 
We find that singlet superconductivity is impossible; instead, triplet superconductivity is favored, with four solutions that are close to each other in energy and differ by their symmetry; one is f-wave, the other three p-wave, including a p$+i$p solution that breaks time-reversal invariance.
\end{abstract}


\maketitle


The electronic properties of graphene have been the object of much research since isolated sheets have been manipulated and characterized in 2004 (for a review, see \cite{Neto:2009ai}).
Graphene has a very high mobility, displays an anomalous quantum Hall effect (at very high field) even at room temperature and also shows a universal conductivity.
The possibility of doping graphene by applying an electric field offers the prospect of carbon-based electronics.
From a theoretical point of view, a peculiar feature of graphene is the cone-like dispersion relation around two unequivalent points in the Brillouin zone, which allows a low-energy description in terms of fermions obeying the $(2+1)$-dimensional Dirac equation.
There have been speculations that a modified graphene system, obtained for instance by stacking planes or by disorder, could display magnetic or even superconducing order at room temperature\cite{Uchoa:2007eu,Honerkamp:2008qd,Pathak:2008qe,Gonzalez:2001if}.
The goal of this work is to check whether electronic correlations alone, as described by a repulsive Hubbard model, can induce superconductivity on doped graphene;
thus we will ignore electron-phonon and long-range Coulomb interactions.
Neglecting the latter is likely a poor approximation very close to half-filling, since screening is then hindered by a small density of states, and we expect our work to be relevant mostly away from that point.

\begin{figure}[tbp]
\centerline{\includegraphics[width=6cm]{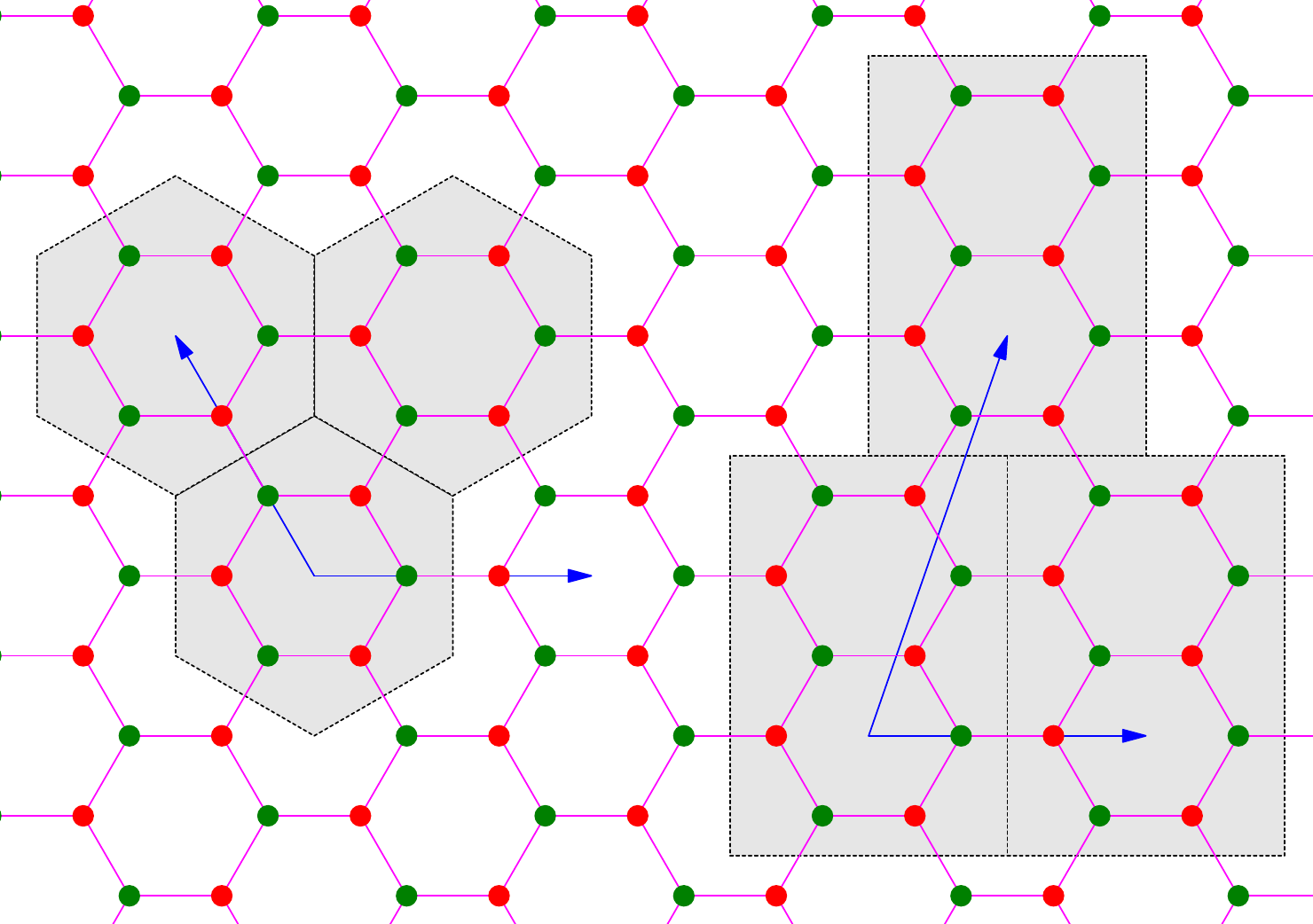}}
\caption{(Color online) 6- and 10-site clusters used in this work. The superlattice vectors are indicated by red arrows. The A and B sublattices are indicated by black and blue dots, respectively.}
\label{fig:hexa}
\end{figure}

Our study focuses on the two-dimensional Hubbard model defined on the honeycomb lattice with nearest-neighbor hopping amplitude $t$, representing interactions within the $\pi$ band:
In pure graphene and three-dimensional graphite, this band is half-filled.
$t$ is estimated at 2.8 eV and the on-site Coulomb repulsion $U$ is expected to be around 17 eV \cite{Neto:2009ai}. 
We will set $U=6t$ in what follows.
The honeycomb lattice is bipartite: it can be viewed as the sum of two interspersed  triangular sublattices (A and B, see Fig.~\protect\ref{fig:hexa}).
The Hubbard model Hamiltonian is expressed as
\BE\label{eq:Hubbard}
H=- t\kern-0.2em\sum_{\rv\in A,\sigma,j}\l( c^\dagger_{\rv,\sigma}c_{\rv+\ev_j,\sigma} + \mathrm{H.c}\r) + U \sum_{\rv\in A, B} n_{\rv,\uparrow} n_{\rv,\downarrow}
\EE
where $c^{(\dagger)}_{\rv,\sigma}$ destroys (creates) and electron of spin $\s$ in a Wannier orbital at site $\rv$, and $n_{\rv,\s} = c^{(\dagger)}_{\rv,\sigma}c_{\rv,\sigma}$ is the number of electrons of spin $\s$ at site $\rv$.
The three vectors $\ev_{1,2,3}$ link a site of sublattice A with its three nearest neighbors (NN) on sublattice B, and are oriented at $120^\circ$ of each other.
The first sum runs over sites of the A sublattice only and contains all hopping terms. 
The second sum (the local Coulomb repulsion) runs over all sites.

Superconductivity is characterized by the condensation of Cooper pairs, which translates into a nonzero expectation value for some pairing operator.
In momentum space, it is convenient to arrange the creation and annihilation operators for the two sublattices and the two spins into a four-component object $f = (c_{A,\kv,\up},c_{B,\kv,\up},c^\dagger_{A,-\kv,\dn},c^\dagger_{B,-\kv,\dn})$.
Pairing may occur either in the singlet or triplet channel; we will show below that the latter is favored.
In the mean-field approximation, the Hamiltonian of the system would take the form $H_{\rm MF} = f^\dagger {\cal H}f$. In the triplet channel, ${\cal H}$ is
\BE
{\cal H} = \begin{pmatrix}
-\mu  & -\g_\kv  & -i\theta_\kv & \eta_\kv \\
-\g_\kv^* & -\mu  & -\eta_\kv^* & -i\theta_\kv \\
i\theta_\kv & -\eta_\kv & \mu  & \g_\kv  \\
\eta_\kv^*  & i\theta_\kv & \g_\kv^* & \mu 
\end{pmatrix}
\EE
where $\mu$ is the chemical potential, $\g_\kv = t\sum_{j=1,2,3} \e^{i\kv\cdot\ev_j}$ is the hopping function and $\theta_\kv$ and $\eta_\kv$ describe pairing amplitudes between electrons of the same sublattice and different sublattices, respectively (this parametrization assumes the pairings and hopping terms to be real when expressed in real space).

It is a straightforward matter to show that the (four-branch) dispersion relation derived from this mean-field Hamiltonian is
\begin{widetext}
\BE\label{eq:dispersion}
E_\kv = \pm\sqrt{|\g_\kv|^2+|\eta_\kv|^2+\theta_\kv^2+\mu^2\pm2\sqrt{\mu^2|\g_\kv|^2 + \theta^2_\kv|\eta_\kv|^2 + 2\mu\theta_\kv\text{Im}\l(\g_\kv\eta_\kv^*\r) + [{\rm Re}(\eta_\kv\g_\kv^*)]^2}}
\EE
\end{widetext}
Owing to the two bands of graphene, this expression does not have the classic BCS shape with an easily identifiable gap function $\Delta_\kv$ (the same goes for the singlet channel dispersion).
The question of gap symmetry may instead be investigated by considering the dominant pairings in real space (see also Fig.~\ref{fig:OP} below).
The repulsive local interaction excludes the possibility of on-site pairing (i.e., $\theta_\kv$ has no constant term).
Pairing operators may be defined between nearest-neighbor (NN) sites ($i=1,2,3$), either in the singlet ($-$ sign) or triplet ($+$ sign) channel:
\BE\label{eq:pairing}\begin{aligned}
S_{i,\rv} &= c_{\rv,\up}c_{\rv+\ev_i,\dn} - c_{\rv,\dn}c_{\rv+\ev_i,\up} \\
T_{i,\rv} &= c_{\rv,\up}c_{\rv+\ev_i,\dn} + c_{\rv,\dn}c_{\rv+\ev_i,\up} 
\end{aligned}\EE
The associated pairing functions are simply $\eta_\kv = \sum_{j=1,2,3} \alpha_j \e^{i\kv\cdot\ev_j}$ and $\theta_\kv=0$, where we introduced pairing amplitudes $\alpha_{1,2,3}$ in the three NN directions.
In Ref.~[\onlinecite{Uchoa:2007eu}], the spin-singlet NN pairing $S_i$ was referred to as `p-wave'; we will refrain from using this terminology since it might be confusing in the context of a point-group symmetry description of the possible pairing states.

The honeycomb lattice is characterized by a $C_{6v}$ symmetry about the center of the hexagons ($C_{6v} \sim D_6$ in two dimensions).
A straightforward analysis in terms of group projection operators shows that the six pairings operators (\ref{eq:pairing}) can be arranged into four different irreducible representations of $C_{6v}$:
\BE\begin{aligned}
\Delta_{A_1} &= \alpha(S_1+S_2+S_3) \\
\Delta_{B_1} &= \alpha(T_1+T_2+T_3) \\
\Delta_{E_1} &= \alpha(T_1-T_2) + \beta(T_2-T_3) \\
\Delta_{E_2} &= \alpha(S_1-S_2) + \beta(S_2-S_3)
\end{aligned}\EE
(the $E_1$ and $E_2$ representations are two-dimensional, and $\alpha$, $\beta$ are constants; the site index $\rv$ suppressed).
At the critical temperature, it is expected from Landau theory that the actual superconducting state fall into one of the above group representations.
As the temperature is lowered, additional symmetry breaking transitions may occur so that the zero-temperature state may be a mixture of the above states
(see also the symmetry analysis of Ref.~\cite{Kuroki:2005cl} in the context of Cobaltates).

The mean-field picture is useful to develop a physical sense of the superconducting state, but we will not use it in computations.
Instead, we use the more powerful variational cluster approximation (VCA), at zero-temperature.
The VCA \cite{Potthoff:2003b} is a variational method based on the electron self-energy, as defined in Potthoff's self-energy functional approach (SFA)\cite{Potthoff:2003}.
The basic idea behind the SFA is to introduce a {\it reference} Hamiltonian $H'$, with the same two-body interaction as the original Hamiltonian $H$, but with a different one-body part, so that $H'$ may be solved exactly (numeri\-cally).
In the VCA, the original lattice is tiled into a superlattice of identical clusters and $H'$ is the cluster Hamiltonian; it differs from the original Hamiltonian $H$ by the suppression of intercluster hopping terms and the addition of Weiss fields associated with the broken symmetry phases of interest.
The values of the Weiss fields, as well as other one-body parameters like the cluster's chemical potential, serve as variational parameters to optimize a functional $\Omega$ whose expression is
\BE\label{eq:omega2}
\Omega({\bf t}')=\Omega'\kern-0.1em - \kern-0.1em\int_C \frac{d\omega}{2\pi}\sum_{\Kv}\ln\det\left(
1 \kern-0.1em + \kern-0.1em (G_0^{-1}\kern-0.2em -G_0'{}^{-1})G'\right)
\EE
where $G_0$ is the non-interacting Green function of the original Hamiltonian, $G_0'$ the non-interacting Green function of the cluster Hamiltonian and $G'$ the exact Green function of the cluster Hamiltonian.
At the physical self-energy, this functional approximates the grand potential $\Omega$ of the system.
The only approximation comes from the limited space of self-energies on which the variational principle is applied, limited by the cluster size and by the number of variational parameters used.

\begin{figure}[t]
\centerline{\includegraphics[width=8.0cm]{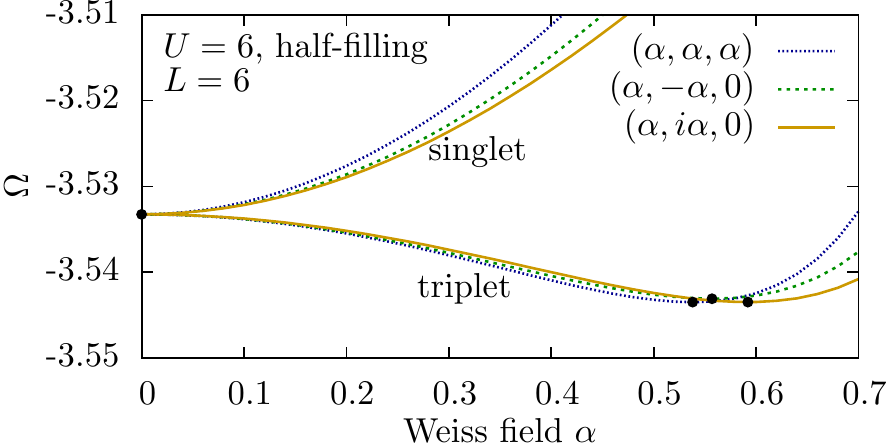}}
\caption{(Color online) Sample Potthoff functional $\Omega$ as a function of Weiss field, for three triplet solutions and the corresponding singlet solutions, at half-filling. The minima are shown as dots. Solution (\ref{eq:E2}) is not shown since it involves two variational parameters.}
\label{fig:omega}
\end{figure}

\begin{figure}[b]
\centerline{\includegraphics[width=8.8cm]{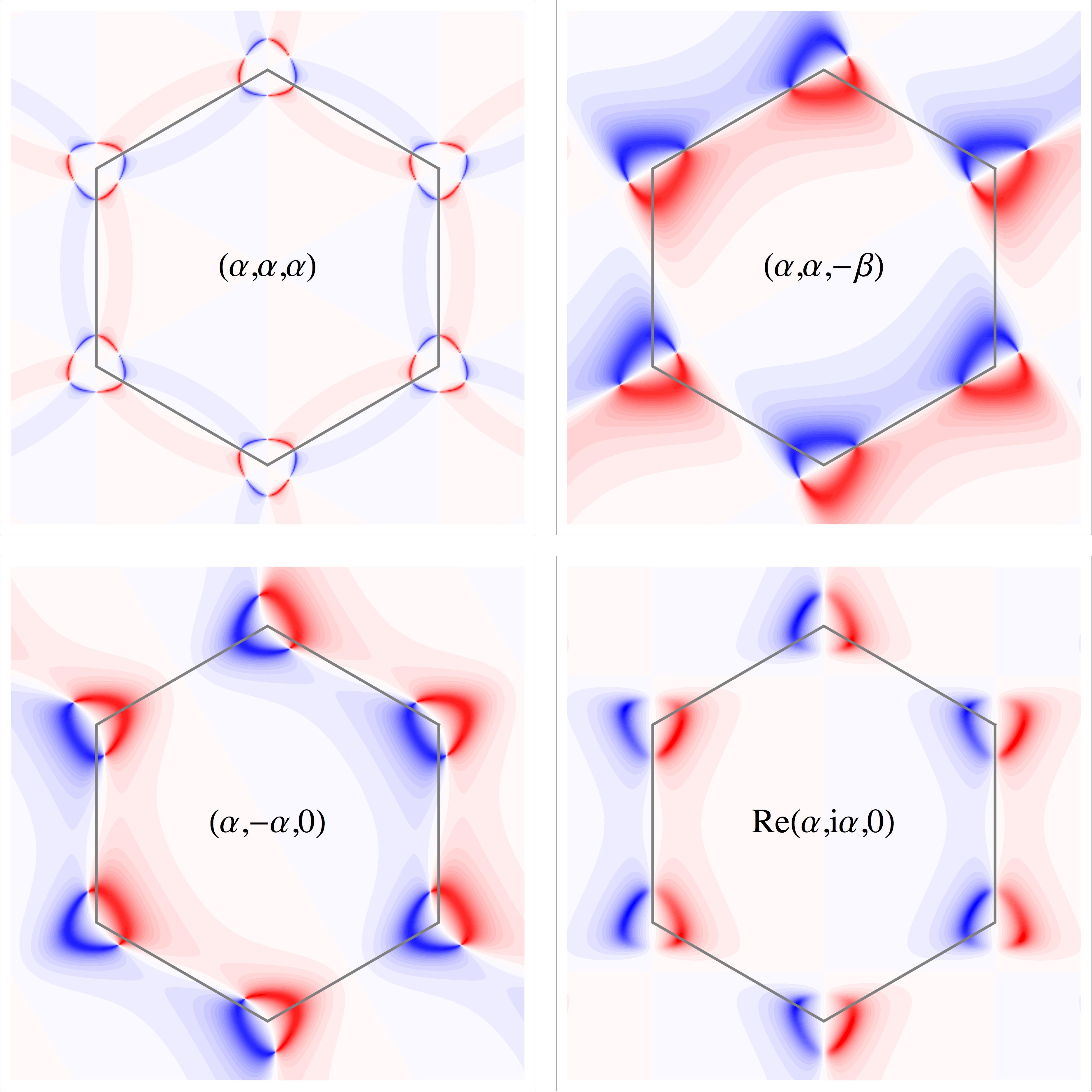}}
\caption{(Color online)
Density plots of the order parameter $\langle c_{A,\kv,\up}c_{A,-\kv,\dn}\rangle$ as a function of $\kv$, for mean-field solutions having the same symmetry as the four triplet solutions found. The Brillouin zone (hexagon) is indicated. Negative (positive) values are represented by shades of blue (red); white is zero.
Note how the topology of nodes (white lines) differs among the four solutions. The bottom-right panel only shows the real part of a complex order parameter; the imaginary part has conjugate nodes, so that the modulus of the order parameter is constant along the Fermi surface.}
\label{fig:OP}
\end{figure}

The VCA has been applied, for instance, to the problem of competing phases in the high-$T_c$ cuprates \cite{Senechal:2005, Aichhorn:2006a} and in the layered organic conductors \cite{Sahebsara:2006bs}.
The VCA does not require any factorization of the interaction, and short-range correlations (within a cluster) are taken into account exactly. 
The Green function obtained from VCA is still defined on the infinite lattice. 

Fig.~\ref{fig:hexa} illustrates the two clusters ($L=6$ and $L=10$ sites, respectively) that were used in this work.
While it is not possible in this case to perform any kind of finite-size scaling analysis, comparing the results from different clusters is useful in order to assess their robustness.
The 6-site cluster has the advantage of posessing the $C_{6v}$ symmetry of the lattice.
The variational parameters used in is this work are the (complex) coefficients of the six different pairing operators (\ref{eq:pairing}), as well as the cluster's chemical potential $\mu'$ (including the latter in the variational set garantees thermodynamic consistency\cite{Aichhorn:2006a}).

Four different superconducting solutions were found, all of them spin triplets.
{\it Thus the first conclusion of this work is that singlet superconductivity does not occur in this system through a purely repulsive interaction.}
Each of the four solutions can be described by the coefficients $(\alpha_1,\alpha_2,\alpha_3)$ of the triplet paring operator $\alpha_1 T_1 + \alpha_2 T_2 + \alpha_3 T_3$:
\begin{align}
&(\alpha,\alpha,\alpha)\qquad && B_1~\text{representation} && (\mathrm{f})\label{eq:E1}\\
&(\alpha,-\alpha,0)&&  E_1~\text{representation} && (\mathrm{p})\label{eq:E3}\\
&(\alpha,\alpha,-\beta) &&B_1-E_1~\text{mixture} && (\mathrm{p})\label{eq:E2}\\
&(\alpha,i\alpha,0) && \text{broken T-reversal} && (\mathrm{p}+i\mathrm{p})\label{eq:E4}
\end{align}
Solutions (\ref{eq:E3}), (\ref{eq:E2}) and (\ref{eq:E4}) also exist in rotated form, obtained for instance by permuting the coefficients.
Fig.~\ref{fig:omega} illustrates the dependence of the Potthoff functional $\Omega$ on the Weiss field $\alpha$ for three of the above solutions, as well as for the corresponding trial singlet-SC states. 
The only extrema of the singlet states occurs at $\alpha=0$, hence the absence of singlet-SC order.
However, the triplet states have nontrivial minima.
It has been argued\cite{Kuroki:2001ye} that, in the presence of disconnected Fermi surfaces on the triangle or honeycomb lattices, triplet pairing may be favored over singlet pairing.

The nodal structure of each of the four solutions (\ref{eq:E1}--\ref{eq:E4}) is illustrated on Fig.~\ref{fig:OP} in plots of the order parameter $\langle c_{A,\kv,\up}c_{A,-\kv,\dn}\rangle$, taken from corresponding mean-field solutions.
Solution (\ref{eq:E1}), which is a pure $B_1$ representation, has three nodal lines around each Dirac point and may be thus qualified as f-wave. The other three all have a $E_1$ component and display a single nodal line: they are p-wave.
The last solution (\ref{eq:E4}) is complex and only its real part is represented; its imaginary part has a similar structure with conjugated nodes and its modulus is constant around each Dirac point; it is a p$+i$p solution.
It can be shown that if only NN pairing is present in the mean-field solution of type (\ref{eq:E1}), then the dispersion relation (\ref{eq:dispersion}) amounts to a simple rescaling of the noninteracting case (a renormalization of the Fermi velocity) and superconductivity is a hidden order.
This is due to the pairing function $\eta_\kv$ being proportional to the hopping function $\g_\kv$ in that case and is an artefact of the restriction to NN pairing.
In reality, pairing extends to further neighbors (even if NN pairing only is used as a Weiss field in VCA).
For this reason, we added a third-neighbor pairing term to the mean-field Hamiltonian of solution (\ref{eq:E1}) in order to produce the top-left plot on Fig.~\ref{fig:OP}.

\begin{figure}
\centerline{\includegraphics[width=8.8cm]{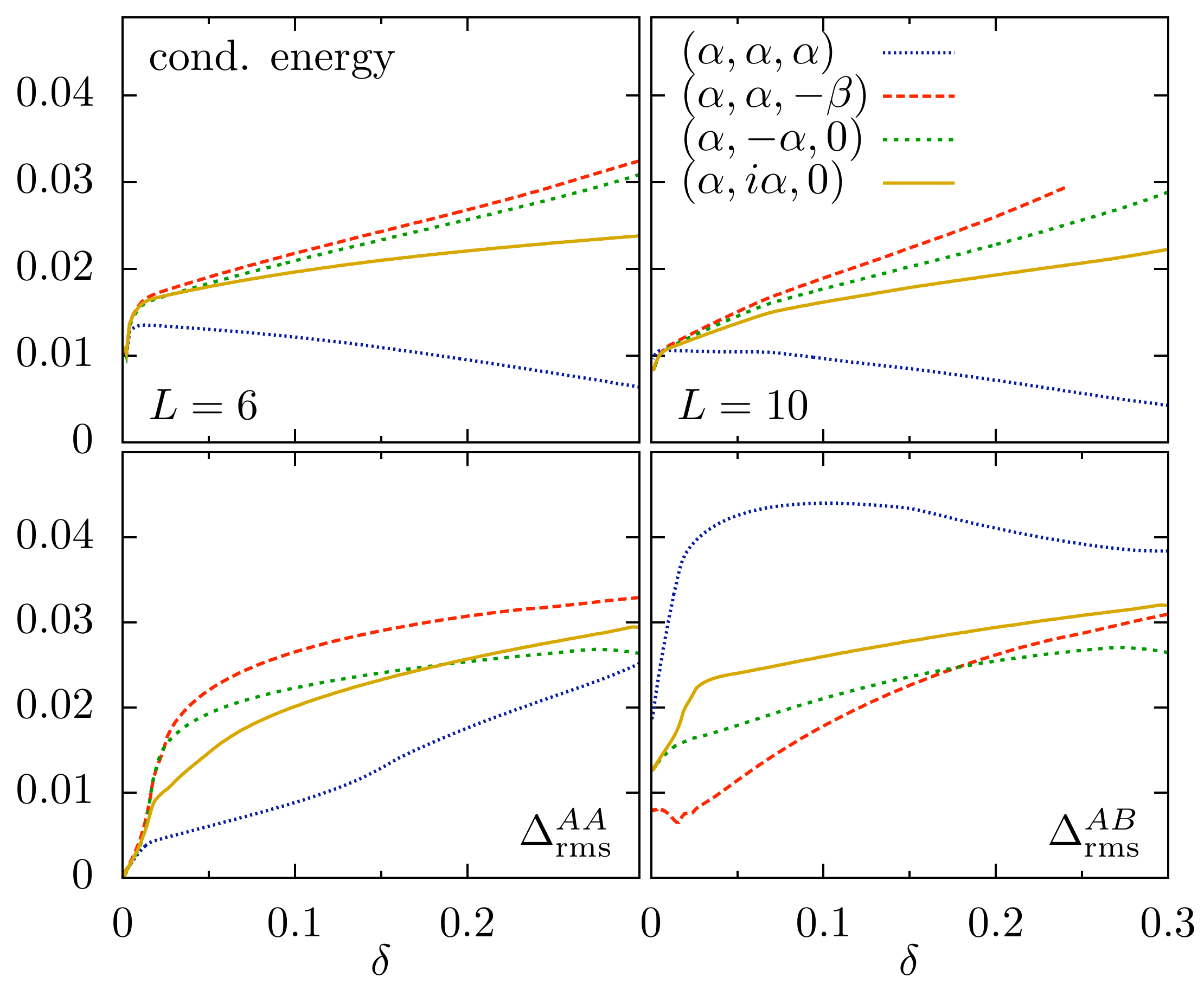}}
\caption{(Color online) Top panels : Condensation energy (in units of $t$) for the various triplet superconducting solutions as a function of hole doping ($\delta = 1-n$), obtained with 6-site (left) and 10-site (right) clusters, at $U=6$. Bottom panels: Root-mean square order parameter for the same sublattice (left) and different sublattices (AB), as a function of doping.}
\label{fig:CE-OP}
\end{figure}

For each value of the chemical potential $\mu$, a normal-state solution can also be found, by suppressing the Weiss fields.
The values of the Potthoff functional for these different solutions gives an estimate of the energy density $E=\Omega + \mu n$, as a function of electron density $n = \Tr G$.
The top panels of Fig.~\ref{fig:CE-OP} show the condensation energy (i.e. the energy of the normal solution, minus that of the ordered solution) for each of the four SC solutions as a function of doping $\delta$.
We see that the p-wave solutions (\ref{eq:E3}), (\ref{eq:E2}) and (\ref{eq:E4}) are close in energy to each other and seem to be favored at larger coupling (except close to half-filling).
As an additional measure of the strength of the SC solutions, we computed the root-mean square (RMS) value $\Delta_\mathrm{rms}^{\mu\nu}$ of the momentum-dependent order parameter:
\BE
\l(\Delta_\mathrm{rms}^{\mu\nu}\r)^2 = \int\frac{\mathrm{d}^2k}{(2\pi)^2} |\langle c_{\mu,\kv,\up}c_{\nu,-\kv,\dn}\rangle|^2
\EE
where $\mu$ and $\nu$ are sublattice indices ($A$ or $B$).
This is illustrated on the bottom panels of Fig.~\ref{fig:CE-OP} for the 6-site cluster.
Note that the order parameter falls towards half-filling, and that the diagonal piece $\Delta_\mathrm{rms}^{AA}$ falls to zero.
The diagonal and off-diagonal parts are similar, except for the f-wave solution (\ref{eq:E1}).
A condensation energy of $0.01t$ corresponds, in this case, to an energy scale roughly equal to room temperature,
but infinite-range order in a two-dimensional system would not exist at a nonzero temperature.

The possibility of ferromagnetic N\'eel order, arising from the bipartite character of the lattice, was also investigated (this order is defined at zero wavevector ($\Qv=0$), hence the term `ferromagnetic').
A Weiss field $M$, multiplying the staggered magnetization operator
\BE
\hat M = \sum_{\rv\in A} (n_{\rv,\up}-n_{\rv,\dn}) - \sum_{\rv\in B} (n_{\rv,\up}-n_{\rv,\dn})
\EE
was also treated as a variational parameter.
No such order was found away from half-filling.
At half-filling, a N\'eel order was found for $U\gtrsim3$, and that solution is energetically favored over the SC solutions for $U\gtrsim6.5$.
As said above, the relevance of the (non-extended) Hubbard model at half-filling in this sytem is unclear.
However, away from half-filling, the exchange of ferromagnetic fluctuations is a possible mechanism of triplet pairing.

{\it Thus, our second conclusion is that triplet superconductivity exists in the doped system, and that a p-wave solution seems favored.}
It is impossible to reliably state which of the three solutions (\ref{eq:E3},\ref{eq:E2},\ref{eq:E4}) is preferred in the thermodynamic limit.
In the case where solution (\ref{eq:E4}) is preferred, time-reversal invariance would be spontaneously broken.

Remains the issue of the physical realization of graphene sheets with sufficient doping for this prediction to be tested.
At present, only very small doping can be achieved by applying electric fields ($\delta\sim 10^{-4}$).
DFT calculations show that coating graphene with a metal can induce a shift in chemical potential of $\sim 0.5$eV \cite{Giovannetti:2008hq}, which translates, in the non-interacting case, into $\delta\sim 0.03$, but which may translate into an even smaller value if correlations are taken into account.
Chemical doping of some kind may be the only possible way to reach values of $\delta$ relevant to this work.

Discussions with A.-M.~S.~Tremblay are gratefully acknowledged.
This work was supported by NSERC (Canada).
Computational resources were provided by the R\'eseau qu\'eb\'ecois de calcul de haute performance (RQCHP).


\end{document}